\documentstyle[12pt]{article}
\input psfig.sty
\def\double12 {\smallskipamount=6pt plus2pt minus2pt
                  \medskipamount=12pt plus4pt minus4pt
                  \bigskipamount=24pt plus8pt minus8pt
                  \normalbaselineskip=24pt plus0pt minus0pt
                  \normallineskip=2pt
                  \normallineskiplimit=0pt
                  \jot=6pt
                  {\def\smallskip {\vskip\smallskipamount}}
                  {\def\medskip   {\vskip\medskipamount}}
                  {\def\bigskip   {\vskip\bigskipamount}}
                  {\setbox\strutbox=\hbox{\vrule
                    height17.0pt depth7.0pt width 0pt}}
                  \parskip 0pt
                  \normalbaselines}

\def\half12 {\smallskipamount=6pt plus2pt minus2pt
                  \medskipamount=12pt plus4pt minus4pt
                  \bigskipamount=24pt plus8pt minus8pt
                  \normalbaselineskip=16pt plus0pt minus0pt
                  \normallineskip=2pt
                  \normallineskiplimit=0pt
                  \jot=6pt
                  {\def\smallskip {\vskip\smallskipamount}}
                  {\def\medskip   {\vskip\medskipamount}}
                  {\def\bigskip   {\vskip\bigskipamount}}
                  {\setbox\strutbox=\hbox{\vrule
                    height17.0pt depth7.0pt width 0pt}}
                  \parskip 0pt
                  \normalbaselines}

\def\single12 {\smallskipamount=3pt plus2pt minus2pt
                  \medskipamount=6pt plus4pt minus4pt
                  \bigskipamount=12pt plus8pt minus8pt
                  \normalbaselineskip=12pt plus0pt minus0pt
                  \normallineskip=1pt
                  \normallineskiplimit=0pt
                  \jot=3pt
                  {\def\smallskip {\vskip\smallskipamount}}
                  {\def\medskip   {\vskip\medskipamount}}
                  {\def\bigskip   {\vskip\bigskipamount}}
                  {\setbox\strutbox=\hbox{\vrule
                    height8.5pt depth3.5pt width 0pt}}
                  \parskip 0pt
                  \normalbaselines}

\def\refitem{\par\noindent\hangindent 20pt}

\def\wisk#1{\ifmmode{#1}\else{$#1$}\fi}

\def\lt     {\wisk{<}}
\def\gt     {\wisk{>}}

\def\lsim   {\wisk{_<\atop^{\sim}}}

\def\amin   {\wisk{^\prime\ }}

\def\deg    {\wisk{^\circ}}

\begin{document}
\pagestyle{plain}

\normalsize
\single12
~		

\vspace{4 cm}
\begin{center}
DIFFUSE MICROWAVE EMISSION SURVEY \\
~  \\	
~  \\  
Al Kogut \\
Hughes STX \\
Code 685 \\
NASA Goddard Space Flight Center \\
Greenbelt, MD 20771 USA
\end{center}

\vspace{2 in}
\noindent
The Diffuse Microwave Emission Survey (DIMES) 
has been selected for a mission concept study
for NASA's New Mission Concepts for Astrophysics program.
DIMES will measure the frequency spectrum of the cosmic microwave
background and diffuse Galactic foregrounds at centimeter wavelengths to
0.1\% precision (0.1 mK), and will map the angular distribution to
20 $\mu$K per 6\deg ~field of view.  
It consists of a set of narrow-band cryogenic radiometers,
each of which compares the signal from the sky
to a full-aperture blackbody calibration target.
All frequency channels compare the sky to the {\it same}
blackbody target, with common offset and calibration,
so that deviations from a blackbody spectral {\it shape}
may be determined with maximum precision.
Measurements of the CMB spectrum complement CMB anisotropy experiments
and provide information on the early universe
unobtainable in any other way;
even a null detection will place important constraints
on the matter content, structure, and evolution of the universe.
Centimeter-wavelength measurements of the diffuse Galactic emission
fill in a crucial wavelength range
and test models of the heat sources, energy balance, and composition
of the interstellar medium.

\clearpage
\normalsize
\half12
\section{Introduction}

The origin and evolution of the 
observed structures in the present-day universe 
is a fundamental unsolved problem of cosmology.
The paradigm for structure formation
consists of the gravitational infall and collapse of
small ``seed'' perturbations in the density of the early universe.
The central result of the {\it Cosmic Background Explorer (COBE)}
was the support of this basic picture
through the detection of CMB anisotropy
without a corresponding distortion from a blackbody spectrum.
Within this broad outline, however, a number of more detailed questions
remain unanswered.
What mechanism created the metric perturbations which serve
as the seeds for structure formation?
What is the nature of ``dark matter'', 
inferred through its gravitational effects?
When and how did the first collapsed objects form?

\single12
\begin{figure}[b]
\centerline{
\psfig{file=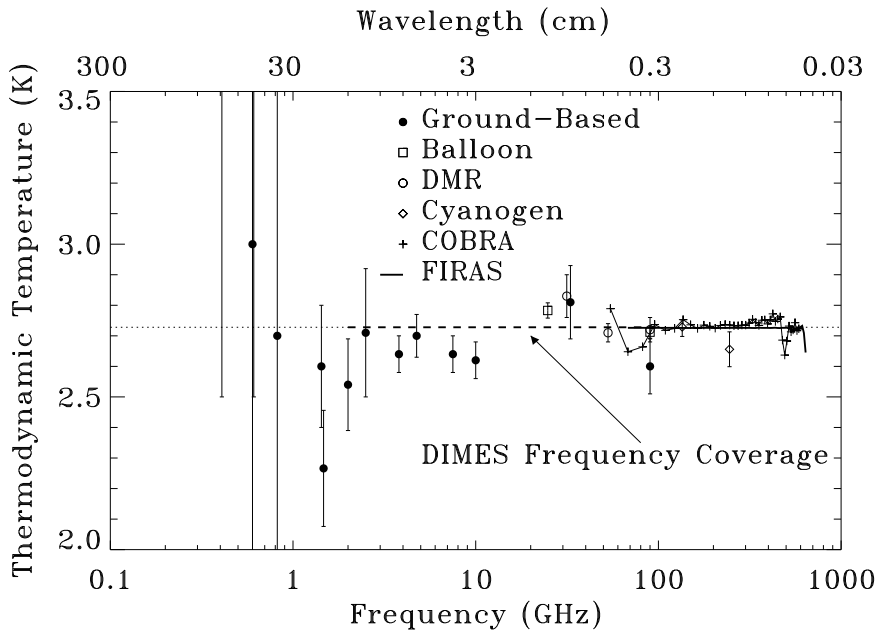,width=4.5in,height=3.0in,angle=90}}
\caption[Recent precise measurements of CMB spectrum]
{\small Precise measurements of the CMB thermodynamic temperature.
The dotted line represents a 2.73 K blackbody,
while the dashed line shows the DIMES frequency band.
The CMB spectrum is poorly constrained at centimeter or longer wavelengths.}
\label{recent_spectrum_fig}
\end{figure}

\half12
To answer these questions, we must turn to relics from the early universe.
The CMB spectrum records the energetics of the universe 
from redshift $z \approx 10^7$ (about one year after the Big Bang)
to the present epoch.
Distortions in the CMB spectrum away from a Planck spectrum
are an unavoidable consequence of structure in the early universe.
Precise measurements of the CMB spectrum at centimeter wavelengths
probe different physical processes than the {\it COBE} results 
at millimeter and sub-mm wavelengths 
and provide quantitative information 
on the energetics of the evolving universe, 
including:

\vspace{2mm}
\refitem
$\bullet$~The photo-ionization of the intergalactic medium
by the first generation of collapsed objects

\vspace{2mm}
\refitem
$\bullet$~The abundance, lifetimes, and decay modes of 
non-baryonic particles, including massive neutrinos,
supersymmetric partners of known particles,
or other dark matter candidates

\vspace{2mm}
\refitem
$\bullet$~The decay of short-wavelength metric perturbations
and primordial turbulence

\vspace{2mm}
\noindent
Figure \ref{recent_spectrum_fig} summarizes recent precise measurements 
of the CMB spectrum.
The {\it COBE} Far Infrared Absolute Spectrophotometer (FIRAS)
provides a precise determination of the CMB spectrum 
near its peak at millimeter and sub-mm wavelengths,
limiting deviations from a blackbody to be
less than $5 \times 10^{-5}$ of the peak intensity$^{1)}$.
Although the precise {\it COBE} measurements carry implications
for possible distortions at longer wavelengths,
the absence of distortions near the peak CMB intensity does {\it not}
imply correspondingly small distortions at longer wavelengths.
Distortions as large as 5\% could exist 
at wavelengths of several centimeters or longer
without violating existing observations$^{2)}$.

Where should further effort, if any, be directed?
Several factors point to spectrum measurements at centimeter wavelengths
(frequencies 2--100 GHz) to precision 0.1 mK as both
technologically feasible and cosmologically interesting.
Measurements in this band probe different physical processes 
than the {\it COBE} measurements a decade higher in frequency,
yet lie at wavelengths short enough to reduce Galactic emission
to manageable levels.  
A multi-channel experiment with 0.1 mK precision per channel
would improve existing cm-band results by a factor of 300,
limit energetic events to $\Delta E/E < 10^{-5}$
at redshift $10^4 < z < 10^7$,
and provide a decisive test of reionization from the first collapsed objects.

\section{Spectral Distortions}
The simplest distortion results from free-free cooling
of a warm plasma (electron temperature $T_e \sim 10^4$ K) 
at recent epochs ($z \lt 1000$),
corresponding to photoionization of the intergalactic medium
by the first generation of stars and galaxies.
The distortion to the present-day CMB spectrum is given by$^{3)}$
\begin{equation}
\Delta T_{\rm ff} = T_{\gamma} \frac{Y_{\rm ff}}{x^2}
\label{FF_distortion_eq}
\end{equation}
where $T_{\gamma}$ is the undistorted photon temperature,
$x$ is the dimensionless frequency $h \nu / k T_{\gamma}$,
$Y_{\rm ff}$ is the optical depth to free-free emission
\begin{equation}
Y_{\rm ff} = \int^z_0 ~\frac{ k[T_e(z) - T_{\gamma}(z)] }{ T_e(z) }
\frac{ 8 \pi e^6 h^2 n_e^2 g }{ 3 m_e (kT_{\gamma})^3 \sqrt{6\pi m_e k T_e} }
\frac{dt}{dz\amin} dz\amin,
\label{Yff_definition}
\end{equation}
$n_e$ is the electron density, and g is the Gaunt factor.
The distorted CMB spectrum is characterized by a quadratic rise
in temperature at long wavelengths as the photon distribution
thermalizes to the plasma temperature.

If the gas is sufficiently hot ($T_e > 10^6$ K),
Compton scattering 
($\gamma + {\rm e} \rightarrow \gamma\amin + {\rm e}\amin$)
of the CMB photons from the hot electrons 
provides the primary cooling mechanism.
Compton scattering transfers energy from the electrons to the photons 
while keeping the photon number fixed.
For recent energy releases ($z \lt 10^5$),
the gas is optically thin,
resulting in a uniform decrement
$ \Delta T_{\rm RJ} = T_{\gamma} (1 - 2y) $
in the Rayleigh-jeans part of the spectrum where there are now too few photons, 
and an exponential rise in temperature in 
the Wien region where there are now too many photons.  
The magnitude of the distortion is related 
to the total energy transfer$^{4)}$
$$
\frac{\Delta {\rm E}}{\rm E} = {\rm e}^{4y} - 1 \approx 4y
$$
where the parameter $y$ is given by the integral 
\begin{equation}
y = \int^z_0 ~\frac{ k[T_e(z) - T_{\gamma}(z)] }{ m_e c^2} 
\sigma_T n_e(z) c \frac{dt}{dz\amin} dz\amin,
\label{compton_y_definition}
\end{equation}
of the electron pressure $n_e k T_e$ along the line of sight 
and $\sigma_T$ denotes the Thomson cross section.

\single12
\begin{figure}[t]
\centerline{
\psfig{file=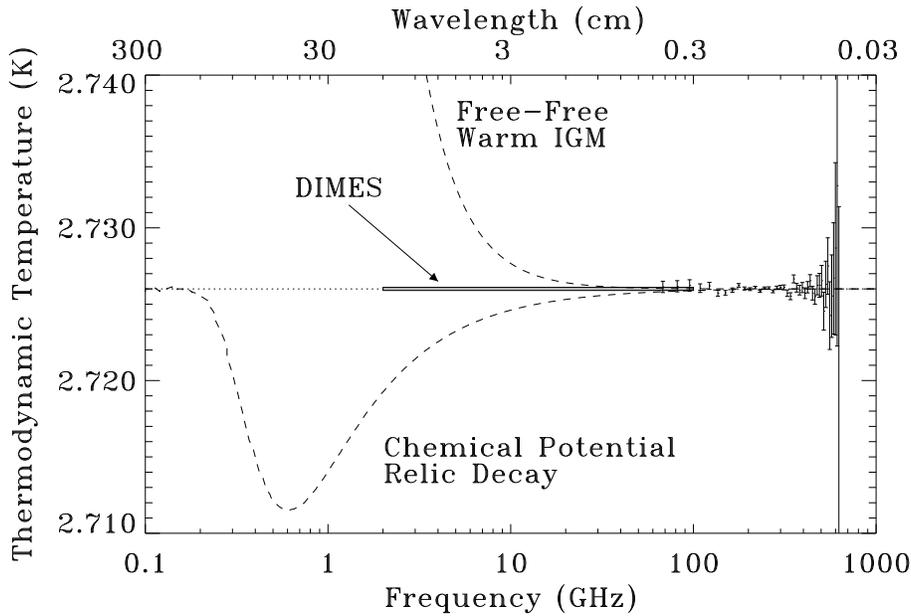,width=4.5in,height=3.0in,angle=90}}
\caption[Upper limits to distorted spectra]
{\small Current 95\% confidence upper limits to distorted CMB spectra.
The FIRAS data and DIMES 0.1 mK error box are also shown;
error bars from existing cm-wavelength measurements are larger than
the figure height.
An absence of distortions at millimeter and sub-mm wavelengths does {\it not}
imply correspondingly small distortions at centimeter wavelengths.}
\label{dimes_vs_firas}
\end{figure}

\half12
Together, free-free and Comptonized spectra
can be used to detect the onset of nuclear fusion 
by the first collapsed objects.
Ultraviolet radiation from the first collapsed objects is expected to
photoionize the intergalactic medium.
Since these objects form by non-linear collapse of rare high-density peaks
in the primordial density distribution,
the redshift at which they form is a sensitive probe of the 
statistical distribution of density peaks 
and the matter content of the universe.
Various models$^{5,6)}$ of structure formation 
predict significant ionization
at redshifts ranging from $10 < z < 150$,
depending on the matter content and power spectrum of density perturbations,
with a ``typical'' value $z_{\rm ion} \approx 50$.

\single12
\begin{figure}[b]
\centerline{
\psfig{file=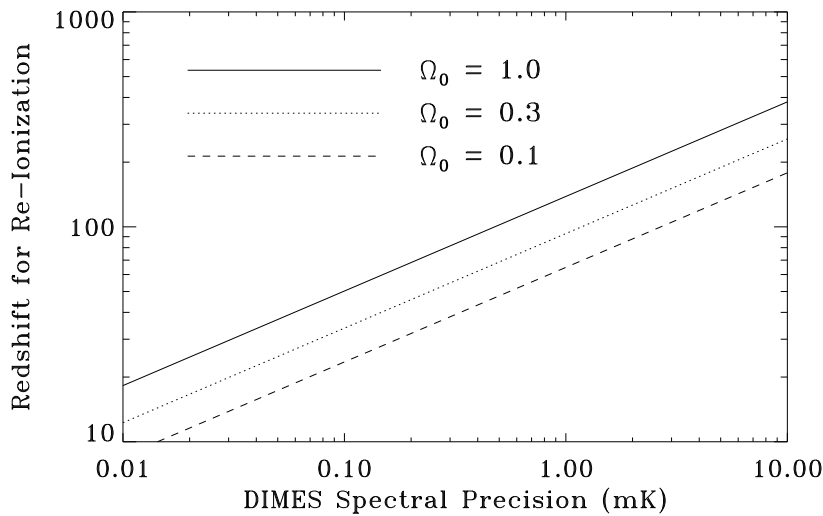,width=4.5in,height=3.0in,angle=90}}
\caption[Redshift for reionized IGM]
{\small Upper limits to the redshift $z_{\rm ion}$ at which the
intergalactic medium becomes reionized, as a function of the
DIMES spectral precision.
The cosmologically interesting region $z_{\rm ion} < 100$
requires precision 0.1 mK or better.}
\label{dimes_reion} 
\end{figure}

\half12
The FIRAS measurement at sub-mm wavelengths shows no evidence
for Compton heating from a hot IGM.
Since the Compton parameter 
$y \propto n_e T_e$ (Eq. \ref{compton_y_definition}),
the IGM at high redshift must not be very hot ($T_e \lsim 10^5$ K)
or reionization must occur relatively recently ($z_{\rm ion} < 10$).
DIMES provides a definitive test of these alternatives.
Since the free-free distortion 
$Y_{\rm ff} \propto n_e^2 / \sqrt{T_e}$ (Eq. \ref{Yff_definition}),
lowering the electron temperature {\it increases} 
the spectral distortion$^{3)}$.
Figure \ref{dimes_reion} shows the limit to $z_{\rm ion}$
that could be established from the combined DIMES and FIRAS spectra,
as a function of the DIMES sensitivity.
A spectral measurement at centimeter wavelengths with 0.1 mK precision 
can detect the free-free signature from the ionized IGM,
allowing direct detection of the onset of hydrogen burning.

DIMES also provides a sensitive test for early energy releases,
such as the decay of exotic heavy particles or metric perturbations
from GUT and Planck-era physics.
Compton scattering from an early energy release 
alters the photon energy distribution while conserving 
photon number.  After many scatterings the system will reach statistical
(not thermodynamic) equilibrium, described by the Bose-Einstein distribution
with dimensionless chemical potential
$ \mu_0 = 1.4 \frac{\Delta {\rm E}}{\rm E}. $
Free-free emission thermalizes the spectrum to the plasma temperature
at long wavelengths.
Including this effect, the chemical potential becomes frequency-dependent,
$$
\mu(x) = \mu_0 \exp(- \frac{2x_b}{x}),
$$
where $x_b$ is the transition frequency at which
Compton scattering of photons to higher frequencies
is balanced by free-free creation of new photons.
The resulting spectrum has a sharp drop in brightness temperature
at centimeter wavelengths$^{7)}$.

DIMES will provide a substantial increase in sensitivity
for non-zero chemical potential (Figure \ref{dimes_vs_firas}).
Such a distortion arises naturally in several models.
The {\it COBE} anisotropy data are well-described$^{8)}$ by a 
Gaussian primordial density field with power spectrum 
$P(k) \propto k^n$ per comoving wave number $k$, 
with power-law index $n = 1.2 \pm 0.3$.
Short-wavelength fluctuations which enter the horizon while the universe is
radiation-dominated oscillate as acoustic waves of constant amplitude and are
damped by photon diffusion, transferring energy from the acoustic waves to the
CMB spectrum and creating a non-zero chemical potential$^{9,10)}$.  
The energy transferred, and hence the magnitude of the present
distortion to the CMB spectrum, depends on the amplitude of the perturbations
as they enter the horizon through the power-law index $n$.
Models with ``tilted'' spectra $n \gt 1$ 
produce observable distortions.

Exotic particle decay provides another source for non-zero chemical potential. 
Particle physics provides a number of dark matter candidates, 
including massive neutrinos, photinos, axions, 
or other weakly interacting massive particles (WIMPs).  
In most of these models, the current dark matter 
consists of the lightest stable member of a family of related
particles, produced by pair creation in the early universe.  Decay of
the heavier, unstable members to a photon or charged particle branch will
distort the CMB spectrum provided the particle lifetime is greater than a year.
Rare decays of quasi-stable particles 
(e.g., a small branching ratio for massive neutrino decay
$\nu_{\rm heavy} \rightarrow \nu_{\rm light} + \gamma$)
provide a continuous energy input, also distorting the CMB spectrum.
The size and wavelength of the CMB distortion are dependent upon the 
decay mass difference, branching ratio, and lifetime. 
Stringent limits on the energy released by exotic particle decay
provides an important input to high-energy theories
including supersymmetry and neutrino physics.

\section{Galactic Astrophysics}
\begin{figure}[b]
\centerline{
\psfig{file=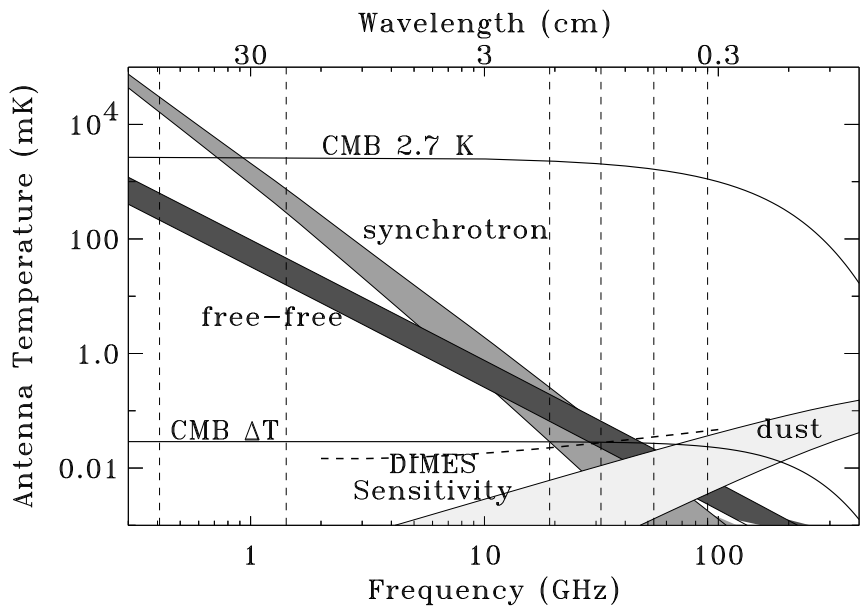,width=4.5in,height=3.0in,angle=90}}
\caption[CMB and Galactic emission spectra]
{\small CMB and Galactic emission spectra.
The shaded regions indicate the signal range 
at high latitude ($|b| > 30\deg$)
and include the effects of spatial structure 
and uncertainties in the spectral index of the Galactic emission components.
Solid lines indicate the mean CMB spectrum and rms amplitude of anisotropy.
Vertical dashed lines indicate existing sky surveys.
The DIMES sensitivity for a 6-month mission is shown.}
\label{foreground_spectra_fig}
\end{figure}

Measurements of the diffuse sky intensity at centimeter wavelengths
also provide valuable information on astrophysical processes within our Galaxy.
Figure \ref{foreground_spectra_fig} shows the relative intensity of 
cosmic and Galactic emission at high galactic latitudes.
Diffuse Galactic emission at centimeter wavelengths
is dominated by three components:
synchrotron radiation from cosmic-ray electrons,
electron-ion bremsstrahlung (free-free emission) 
from the warm ionized interstellar medium (WIM),
and thermal radiation from interstellar dust.
Despite surveys carried out over many years,
relatively little is known about 
the physical conditions responsible for these diffuse emissions.
Precise measurements of the diffuse sky intensity
over a large fraction of the sky,
calibrated to a common standard,
will provide answers to outstanding questions on physical conditions
in the interstellar medium (ISM):

\vspace{2mm}
\refitem
$\bullet$~What is the heating mechanism in the ISM?  Is the diffuse gas 
heated by photoionization from the stellar disk, 
shocks, Galactic fountain flows,
or decaying halo dark matter?  

\vspace{2mm}
\refitem
$\bullet$~How are cosmic rays accelerated?
Is the energy spectrum of local cosmic-ray electrons 
representative of the Galaxy as a whole?

\vspace{2mm}
\refitem
$\bullet$~What is the shape, constitution, and size distribution of 
interstellar dust?  Is there a distinct ``cold'' component in the cirrus?

\single12

\half12
\vspace{2mm}
\noindent
The Galactic radio foregrounds
may be separated from the CMB by their frequency dependence
and spatial morphology.
DIMES will map radio free-free emission from 
the warm ionized interstellar medium.
The ratio of radio free-free emission to H$\alpha$ emission 
will map the temperature of the WIM to 20\% precision,
probing the heating mechanism in the diffuse ionized gas.
DIMES will have sufficient sensitivity to map the high-latitude
synchrotron emission,
probing the magnetic field and electron energy spectrum throughout the Galaxy.
Cross-correlation with the DIRBE far-infrared dust maps
will fix the spectral index of the high-latitude cirrus
to determine whether the dust has
enhanced microwave emissivity.

\section{Instrument Description}
Figure \ref{instrument_schematic} shows a schematic of the DIMES instrument.
It consists of a set of narrow-band cryogenic radiometers
($\Delta \nu / \nu \sim 10\%$)
with central frequencies chosen to cover
the gap between full-sky surveys at radio frequencies ($\nu < 2$ GHz)
and the {\it COBE} millimeter and sub-mm measurements.
Each radiometer measures the difference in power
between a beam-defining antenna (FWHM $\sim 6\deg$)
and a temperature-controlled internal reference load.
An independently controlled blackbody target is located on the aperture plane,
so that each antenna alternately views the sky 
or a known blackbody.
The target temperature will be adjusted 
to null the sky-antenna signal difference
in the longest wavelength channel.
With temperature held constant,
the target will then move to cover the short-wavelength antennas:
DIMES will measure small spectral shifts about a precise blackbody,
greatly reducing dependence on instrument calibration and stability.
The target, antennas, and radiometer front-end amplifiers
are maintained near thermal equilibrium with the CMB,
greatly reducing thermal gradients and drifts.

\begin{figure}[b]
\vspace{3 in}
\caption[DIMES concept]
{\small Schematic drawing of DIMES instrument.}
\label{instrument_schematic}
\end{figure}

DIMES uses multiple levels of differences to reduce the effects of
offset, drifts, and instrumental signatures.
To reduce gain instability or drifts,
each receiver is rapidly switched 
between a cryogenic antenna
and a temperature-controlled internal reference load.
To eliminate the instrumental signature, 
each antenna alternately views the sky or a full-aperture target
with emissivity $\epsilon > 0.9999$.
To maximize sensitivity to spectral shape,
all frequency channels view the {\it same} target in progression,
so that deviations from a blackbody spectrum
may be determined much more precisely 
than the absolute blackbody temperature.

DIMES will remove the residual instrument signature
by comparing the sky to an external full-aperture blackbody target.
The precision achieved will likely to be dominated by
the thermal stability of the target.
While the use of a single external target 
rejects common-mode uncertainties in the absolute target temperature, 
thermal gradients within the target or
variations of target temperature with time
will appear as artifacts in the derived spectra and sky maps.
We reduce thermal gradients within the external target
by using a passive multiply-buffered design
in which a blackbody absorber (Eccosorb CR-112, an iron-loaded epoxy)
is mounted on a series of thermally conductive plates with conductance $G_1$
separated by thermal insulators of conductance $G_2$.
Thermal control is achieved by heating the outermost buffer plate,
which is in weak thermal contact with a superfluid helium reservoir.
Radial thermal gradients at each stage 
are reduced by the ratio $G_2 / G_1$ between the buffer plates.
Typical materials (Fiberglass and copper) achieve 
a ratio $G_2 / G_1 \lt 10^{-3}$;
a two-stage design should achieve net thermal gradients 
well below 0.1 mK.
No heat is applied directly to the absorber,
and a conductive copper layer surrounds the absorber
on all sides except the front:
the Eccosorb lies at the end of an open thermal circuit,
eliminating thermal gradients from heat flow.

DIMES will not be limited by raw sensitivity.
HEMT amplifiers cooled to 2.7 K
easily achieve {\it rms} noise 1 mK Hz$^{-1/2}$,
reaching 0.1~mK sensitivity
in 100 seconds of integration.
The DIMES spectra are derived from comparison of the sky 
to the external blackbody target.
The largest systematic uncertainties arise from
thermal drifts or gradients within the target
and emission from warm objects outside the DIMES dewar (e.g., the Earth).
Thermometers buried in the microwave absorber 
monitor thermal gradients and drifts to precision 0.05 mK.
Emission from the Earth must be rejected at the -70 dB level to
avoid contributing more than 0.1 mK to the total sky signal.
DIMES will achieve this rejection using corrugated antennas with
6\deg ~beam and good sidelobe response;
two sets of shields between the aperture plane and the Earth
provide further attenuation of thermal radiation from the Earth.
{\it COBE} achieved -70 dB attenuation
with a 7\deg ~beam and a single shield$^{11)}$,
so the DIMES requirement should be attainable.

DIMES will eliminate atmospheric emission completely
by observing from low Earth orbit.
We are currently investigating the possibility
of utilizing the Spartan-400 carrier,
which will provide free-flyer capability 
to Shuttle orbits for 700 kg instruments
for a nominal mission of 6 to 9 months.

\normalsize
\single12
\medskip
\begin{center}
{\sc References}
\end{center}
\medskip

\refitem
1) Fixsen, D.J, Cheng, E.S., Gales, J.M., Mather, J.C., Shafer, R.A.,
\& Wright, E.L. 1996, ApJ, submitted

\refitem
2) Bersanelli, M., Bensadoun, M., De Amici, G., Levin, S., Limon, M.,
Smoot, G.F., \& Vinje, W. 1994, ApJ, 424, 517

\refitem
3) Bartlett, J.G., and Stebbins, A., 1991, ApJ, 371, 8

\refitem
4) Sunyaev, R.A., and Zel'dovich, Ya.B., 1970, Ap. Space Sci., 7, 20

\refitem
5) Tegmark, M., Silk, J., and Blanchard, A., 1994, ApJ, 420, 484

\refitem
6) Liddle, A.R. \& Lyth, D.H. 1995, MNRAS, 273, 1177

\refitem
7) Burigana, C., Danese, L., and De Zotti, G.F., 1991, A\&A, 246, 49

\refitem
8) G\'{o}rski K.M., Banday, A.J., Bennett, C.L., Hinshaw, G.,
Kogut, A., Smoot, G.F., \& Wright, E.L., 1996, ApJ, 464, in press

\refitem
9) Daly, R., 1992, ApJ, 371, 14

\refitem
10) Hu, W., Scott, D., and Silk, J., 1994, ApJ., 430, L5

\refitem
11) Kogut, A., et al.\ 1996, ApJ, 470, in press

\end{document}